\documentclass[preprint,showpacs,aps]{revtex4}
\usepackage{mathrsfs}

\usepackage{epsfig}

\begin{document}

\title{\large \bf Domain-wall dynamics of magnetic systems with disorder at zero temperature}

\author{\bf N. J. Zhou$^{1}$, B. Zheng$^{1}$\footnote{corresponding author: zheng@zimp.zju.edu.cn}
and Y.Y. He$^{1}$ }

\affiliation{$^1$ Zhejiang University, Zhejiang Institute of Modern
                  Physics, Hangzhou 310027, P.R. China }

\begin{abstract}
With Monte Carlo simulations, we investigate the relaxation dynamics
with a domain interface in a magnetic system at zero temperature,
taking two-dimensional driven random field Ising model with quenched
disorder as an example. The dynamic scaling behavior is carefully
analyzed, and a dynamic roughening process is observed at the
depinning transition. For comparison, additional simulation without
disorder has been performed as its background. The effect of the
overhangs is discussed, and the growing interface exhibits intrinsic
anomalous scaling and spatial multiscaling with $\zeta > 1$ and
$\zeta_{loc} < 1$, different from the work described by QEW.
\end{abstract}


\maketitle

\section{Introduction}
  In the past years much progress has been achieved in the study of
dynamic processes far from equilibrium. For example, the universal
dynamic scaling form in critical dynamics has been explored up to
the {\it macroscopic} short-time regime
\cite{jan89,luo98,zhe98,zhe99}, when the system is still far from
equilibrium. Although the spatial correlation length is still short
in the beginning of the time evolution, the dynamic scaling form is
induced by the divergent correlating time around a continuous phase
transition. Based on the short-time dynamic scaling form, new
methods for the determination of both dynamic and static critical
exponents as well as the critical temperature have been developed
\cite{li95,luo98,zhe98,zhe99}. Since the measurements are carried
out in the short-time regime, one does not suffer from critical
slowing down. Recent progress in the short-time critical dynamics
includes, for example, theoretical calculations and numerical
simulations of the XY models and Josephson junction arrays
\cite{zhe03a}, magnets with quenched disorder
\cite{yin04,yin05,zho07,zho08}, ageing phenomena \cite{lei07}, weak
first-order phase transitions \cite{sch00,yin05}, and various
applications and developments \cite{lin08}.

  Meanwhile, the physics of elastic systems in disordered media also has
been the focus of intense theoretical and experimental studies in
the different areas, such as charge density waves \cite{lem99},
vortex lattices \cite{fuc98}, domain walls in magnetic \cite{dou08,
yam07} or ferroelectric \cite{tyb02, par05} materials, contact lines
\cite{mou04,dou06}, fluid invasion of porous media \cite{he92},
dislocations moving \cite{zap01,bak08}, and crack propagation
\cite{pon06}. With increasing driving force $F$, the driven
interface in quenched disordered system displays a transition from a
pinned interface to a moving interface. This so-called
pinning-depinning phase transition at zero temperature is viewed as
a critical phenomenon
\cite{rot99,rot01,now98,dou02,jos96,jos97,due05,kol06,kol06a,bus08},
and its ordered parameter is the interface velocity $v$. With the
assumption that the interface shows the properties of an elastic
membrane, this growth process can be described by the driven
Edwards-Wilkinson equation with quenched disordered (QEW) in the
isotropic universality class
\cite{nat92,due05,kol06,jen95,les93,kol06a,bus08} or the driven
quenched Kardar-Parisi-Zhang equation (QKPZ)
\cite{tan92,ros01,ros03,goo04} in the anisotropic universality
class. With these equations, all the critical exponents can be
derived by simulation and theory analysis called functional
renormalization group (FRG) \cite{nat92,dou02,cha00}. And the
roughness exponent $\zeta > 1$ obtained from the equation QEW in $1
+ 1$ dimensions means that the elastic string will inevitably break
as the length of the string is increased \cite{ros01, jen95,
les93,kol06a}.

Recently, these kinetic roughening phenomenons have brought about
interest on the superrough growth of the interface or surface with
$\zeta >1$, not only in simulations \cite{pan00,pan00a,pra08}, but
also in experiment \cite{aug06,sai02,sai03}. Three scaling behaviors
have been observed experimentally, namely, (i) FV standard scaling
for $\zeta = \zeta_{loc} = \zeta_s < 1$; (ii) intrinsic anomalous
scaling for $\zeta \neq \zeta_{loc} = \zeta_s \leq 1$; and (iii)
superrough scaling for $\zeta = \zeta_s > \zeta_{loc} = 1$. Here
roughness exponent $\zeta$, local roughness exponent $\zeta_{loc}$
and spectral roughness exponent $\zeta_s$ can be measured in
different ways, such as globe width function $W(L,t)$, local width
function $W(l,t)$, height correlation function $C(r,t)$, local
interfacial orientation $W_n(l,t)$, and power spectral density
$S(k,t)$. And spatial multiscaling behavior is observed associated
with the intrinsic anomalous scaling class (ii)
\cite{pan00,aug06,pan00a}.

On the other hand, many recent activities have been devoted to the
domain-wall dynamics in the magnetic materials, which is an
important topic in magnetic devices, nano-materials, thin film, and
semiconductors \cite{yam07,dou08,lem98,met07,kle07,bra05,jos98}. And
external field $H$ or electrical current $J$ is applied as the
driving force $F$. At zero temperature, the depinning transition is
obtain when the homogeneous driving field $H$ reaches its critical
value $H_c$, which is relation with the disorder. And if a periodic
external field $H(t)=H_0\cos(\omega t)$ is applied and/or a non-zero
temperature is introduced, the domain wall exhibits different states
of motion and dynamic phase transitions
\cite{nat01,gla03,bra05,kle07}. Most these works concentrate on the
stationary state at the zero or low temperatures and in response to
the external magnetic field $H(t)$. Instead of the phenomenological
model, such as QEW or QKPZ equation, a well established model to
investigate the dynamics behavior of the domain interface is the
driven random field Ising model (RFIM)
\cite{bru84,now98,rot01,rot99,zhe01,sep98}. The scaling behavior and
critical exponents have been obtained for various dimensions at zero
or low temperature, which is roughly consistent with the result of
QEW. While as we well known, the roughness growth is rarely to be
investigated because lots of overhangs occur at the depinning
transition and it is hard to determine the critical exponents and
transition point precisely by the finite-size scaling and
finite-temperature scaling analysis in the stead state

       Very recently, the experiment on the non-stationary dynamic
evolution of the domain wall is reported for the creep and pinning
effect \cite{rod07}, though its dynamic behavior is described
roughly. What's more, the short-time relaxation of a driven elastic
string in a two-dimensional pinning landscape is also reported in
the reference \cite{kol06}, in order to get the critical exponents
of the depinning transition independently. Additional, the short
time dynamic relaxation of a domain wall has been concerned for
magnetic systems at a standard order-disorder phase transition in
the Ising model \cite{zho07,zho08,lin08}. It is described by the
relaxation dynamics starting from a {\it semi-ordered} state, and
shares certain common features with those around free and disordered
surfaces. Since no external magnetic fields are added,
macroscopically the domain wall does not move, but a kind of
roughening phenomena occurs.

   In this paper, we investigate the relaxation dynamics of domain
walls at zero temperature for the depinning transition, taking the
two-dimensions driven RFIM as an example. And we aim at
understanding the non-stationary properties of this dynamics system
and determining the static and dynamic exponents as well as the
transition point. Then the effect of overhangs will be studied by
the roughness scaling behavior, which causes the main difference
between QEW and RFIM at the depinning transition. For comparison,
its background in our work is also investigated without overhangs.
In Sec. II, the model and scaling analysis are described, and in
Sec. III, the numerical results are presented. Sec. IV includes the
conclusions.

\section{Model and scale analysis}
  The random-field Ising model is defined by the Hamiltonian
\begin{equation}
\mathscr{H} = - J \sum_{<ij>}S_iS_j - H \sum_i S_i - \sum_i h_i S_i.
\label{equ10}
\end{equation}
Where $S_i = \pm 1$ is just the Ising spin on the two-dimensional
lattice with $L \times L$. The random field $h_i$ is taken from a
distribution within an interval $[-\Delta, \Delta]$ and hence is
bounded. $H$ is the homogeneous driving field. The strength of the
random field is fixed as $\Delta=1.5 J$ in this paper, and $\Delta =
0 J$ is also used in comparison as its background. For convenience,
we set $J = 1$ in this paper. The main simulation is performed with
the lattice of size $L = 512$ at zero temperature, and the maximum
updating time is $t_M = 1024$. Additional simulation with $L = 256$
and $L = 1024$ are also performed to exclude the finite-size effect.
The total samples for average are $50 000$, and the statistical
error are estimated by dividing the samples into two or three
groups. If the fluctuation in the time direction is comparable with
or larger than the statistical error, it will be taken into account.

   The initial semi-ordered state is such a state that spins are
positive in the sublattice on the left side $x < 20$ and negative on
the right side $x > 20$. Here we set the $x$ axis in the direction
perpendicular the initial interface and its left boundary is located
at $x =0$. Then all the system is rotated by an angle $\pi / 4$ in
order to make sure that the initially interface of the system is
just in the $(11)$-direction of the lattice. Therefore there is no
pinning of the interface without the disorder
\cite{now98,rot99,rot01}. The antiperiodic conditions are used in
the $x$ direction. While in the other direction, periodic conditions
are used. After preparing that, we perform Monte Carlo simulations
and the spins are chosen randomly to update. And the spins will be
flipped only when the total energy decreases after flipped. As time
evolves, the interface moves due to the driving field and becomes
more and more roughening. In Fig. \ref{f1}(a), the real snapshot
images of this interface are illustrated in the dynamic relaxation
process for the depinning transition and its background. The local
interface at the transition point not only tilts, but also forms
local grooves or spikes, quit different from the instance of the
background. Hence, it is difficult to define the height function
$h(t)$ of the interface because of visible overhangs.

   In this case, we define the velocity of the domain interface as the time
derivation of the magnetization\cite{now98}. First, the
magnetization and its second moment are measured,
\begin{equation}
M^{(k)}(t) = \frac{1}{L^{2k}} \left\langle \left[ \sum_{x,y=1}^L
S_{xy}(t) \right]^k\right\rangle, \quad k = 1, 2. \label{equ20}
\end{equation}
Here $S_{xy}(t)$ is a spin at the time t on the lattice position
$(x,y)$, and $<\cdots>$ represents the statistical average over
disorder. For convenience, we also use $M(t) \equiv M^{(1)}(t)$ to
denote the magnetization. Then the velocity of the domain interface
is shown,
\begin{equation}
v(t) = \frac{L}{2}\frac{d \hspace{5pt} M(t)}{d \hspace{5pt} t}.
\label{equ30}
\end{equation}
Here $L / 2$ is the scale factor. In order to characterize the
growth of the domain interface and its fluctuation in the $x$
direction, we introduce a height function and its second moment
\begin{equation}
h^{(k)}(t) = \frac{1}{L^{k}} \left\langle \left[ \sum_{x=1}^L
S_{xy}(t) \right]^k\right\rangle, \quad k = 1, 2. \label{equ40}
\end{equation}
Here $<\cdots>$ represents not only the statistic average but also
the average in the $y$ direction. As usual, we also use the notation
$h(t) \equiv h^{(1)}(t)$ and the relation $h(t) \equiv M(t)$ is
obtained. Then the roughness function of the domain interface is
defined,
\begin{equation}
\omega^2(t) = h^{(2)}(t) - h(t)h(t). \label{equ50}
\end{equation}
After that, the relation between the magnetization fluctuation and
interface fluctuation can be described in the function $F(t)$,
\begin{equation}
F(t) = \frac{M^{(2)}(t) - M(t)M(t)}{h^{(2)}(t) - h(t)h(t)}.
\label{equ60}
\end{equation}
What's more, a more informative quantities for the interface, called
the height correlation function $C(r,t)$, is introduced.
\begin{equation}
C(r, t) = \langle[h(y + r, t) - h(y, t)]^2\rangle. \label{equ70}
\end{equation}
Here $h(y, t)$ describes the local interface height and can be
obtained by Eq.~(\ref{equ40}) with a fixed $y$.

   At the depinning transition point $H = H_c(\Delta)$, the main scale length
in the dynamic system is the non-equilibrium spatial correlation
length $\xi(t)$. For a finite size, the lattice size $L$ is an
additional length scale. And the system exhibits as the second-order
transition when temperature is zero. So one can believe that
$\xi(t)$ grows as a power law $\xi(t) \sim t^{1/z}$ and $z$ is
so-called dynamic exponent. General scaling arguments lead to the
scaling form of the order parameter $v(t)$.
\begin{equation}
v(t) = (\xi(t))^{-\beta/\nu} G(\xi(t) / L, \tau / \xi(t)).
\label{equ80}
\end{equation}
Here $\beta$ and $\nu$ are the static exponents and $\tau = H - H_c$
is defined as the departure from the transition point. And the
scaling function $G(\xi(t) / L, \tau / \xi(t))$ is constant if $L
\rightarrow \infty$ and $\tau = 0$. Then the scaling form is
simplified to
\begin{equation}
v(t) \sim t^{-\beta / \nu z}. \label{equ90}
\end{equation}
And the scaling behavior of the derivative $\partial_{\tau}\ln v(t,
\tau)|_{\tau=0}$ can be obtained according to Eq.~(\ref{equ80}).
\begin{equation}
\partial_{\tau}\ln v(t,\tau)|_{\tau=0} \sim t^{1 / \nu z}.
\label{equ100}
\end{equation}

   In general, the height function $h(t)$, the roughness function
$\omega^2(t)$ and the height correlation function $C(r,t)$ in
Eqs.~(\ref{equ40}), (\ref{equ50}) and (\ref{equ70}) do not obey a
simple power law behaviors. Since the depinning transition is a
dynamic transition and its order parameter is $v(t)\sim dh(t) / dt$.
And in the background without disorder, the interface is also
roughening. So pure roughness function and height correlation
function should be defined by subtracting the contribution from the
background.
\begin{equation}
D\omega^2(t) = \omega^2(t) - \omega_b^2(t). \label{equ110}
\end{equation}
\begin{equation}
DC(r,t) = C(r,t) - C_b(r,t). \label{equ120}
\end{equation}
And the scaling form of $D\omega^2(t)$ and $DC(r,t)$ are shown as in
the case of a standard growing interface, reported in Refs.
\cite{jos96, jos97, pan00}.
\begin{equation}
D\omega^2(t) \sim t^{-2\zeta / z}. \label{equ130}
\end{equation}
\begin{equation}
   DC(r,t) \sim  \left\{
   \begin{array}{lll}
    \xi(t)^{2(\zeta - \zeta_{loc})}r^{2\zeta_{loc}}    & \quad &  \mbox{if  $r \ll \xi(t)$} \\
    \xi(t)^{2\zeta}  & \quad &  \mbox{if  $\xi(t) \ll r$}
   \end{array}\right. .
   \label{equ140}
\end{equation}
Here $\xi(t) \sim t^{1/z}$ and $z$ have been defined before. And
$\zeta$ is the roughness exponent and $\zeta_{loc}$ is the local
one. Since the region with the power law behavior for $DC(r,t)$ vs.
$r$ is rather narrow, a new scaling form is introduced to fit this
relation \cite{jos96,jos97}.
\begin{equation}
DC(r,\infty) = A[tanh(r/B)]^{2\zeta_{loc}}. \label{equ150}
\end{equation}
Then, the roughness function $\omega^2(t)$ represents the
fluctuation only in the $x$ direction, while the fluctuation of the
magnetization is measured in the whole lattice. So the scaling
behavior of the function $F(t)$ is shown,
\begin{equation}
F(t) = \xi(t)^{d - 1} \sim t^{1 / z}. \label{equ160}
\end{equation}

   What's more, all the scaling behaviors above hold only after a time
scale $t_{mic}$, which is sufficiently long in the microscopic
sense, but still short in the macroscopic sense. Hence, a power law
correction to the scaling is used sometimes in order to fit the data
well.
\begin{equation}
Y(x) \sim x^{a}(1 + c / x ). \label{equ170}
\end{equation}
Here the convergence of  $Y(t)$ is to a power law behavior and the
parameter $a$ is viewed as the critical exponent.

\section{The result of numerical simulation}
\begin{figure}[ht]
\epsfysize=7.0cm \epsfclipoff \fboxsep=0pt
\setlength{\unitlength}{1.cm}
\begin{picture}(10,6)(0,0)
\put(-3.0,-0.3){{\epsffile{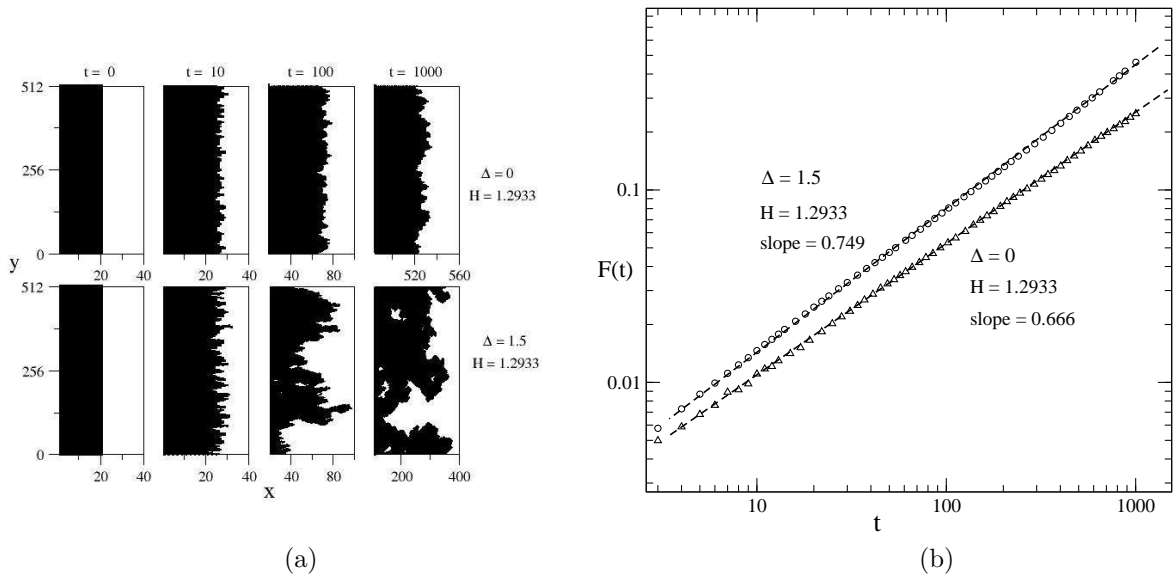}}}\epsfysize=7.0cm
\put(5.2,-0.3){{\epsffile{f_t.eps}}}
\end{picture}

\hspace{1.0cm}\footnotesize{(a)}\hspace{8.0cm}\footnotesize{(b)}
\caption{(a) The time evolution of the interface velocity $v(t)$ is
displayed for different driving field $H$ when the disorder is fixed
as $\Delta = 1.5$. For clarity, the curve with depinning field $H_c
= 1.2933$ is shifted down by a factor $0.5$. Dashed line shows the
power law fit. Open squares corresponds to the different lattice
size with $L = 1024$. \quad(b) The function $F(t)$ is displayed with
open dots on a log-log scale for the depinning transition (above)
and its background (below). And dashed lines show the power law
fits. }\label{f1}
\end{figure}

\begin{figure}[ht]
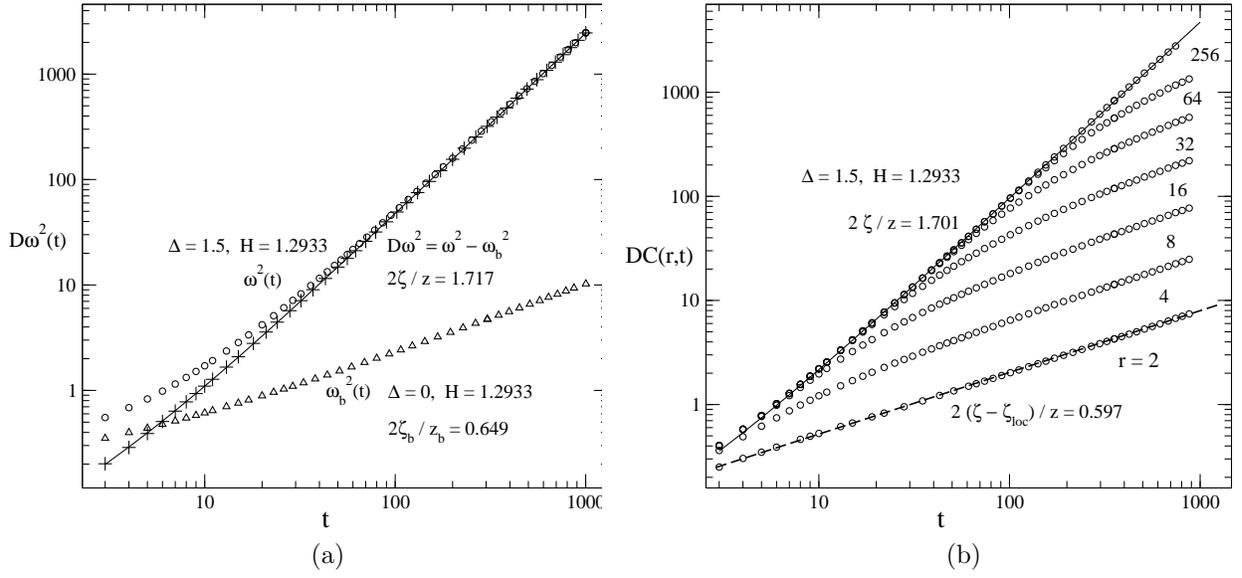

\epsfysize=7.0cm \epsfclipoff \fboxsep=0pt
\setlength{\unitlength}{1.cm}
\begin{picture}(10,6)(0,0)
\put(-3.0,-0.3){{\epsffile{w_1.eps}}}\epsfysize=7.0cm
\put(5.2,-0.3){{\epsffile{c_1.eps}}}
\end{picture}

\hspace{1.0cm}\footnotesize{(a)}\hspace{8.0cm}\footnotesize{(b)}
\caption{(a) The roughness function $\omega^2(t)$ is plotted on a
double-log scale with open dots. And pluses represent the pure
roughness function $D\omega^2(t)$. Solid lines represent power law
fits with correction, basing on Eq.~\ref{equ170}. \quad(b) The pure
height correlation function $DC(r,t)$ is plotted with open circles
on a log-log scale, for $r = 2, 4, 8, 16, 32, 64$ and $256$ (from
below). Dashed lines show the power law fits , and solid line
represents a power law fit with correction. In the inset,
$DC(r,\infty)$ vs. $r$ is plotted for the depinning transition at
its maximum time $t = 1024$. Data are fitted with solid lines,
according to Eq.~\ref{equ150}.}\label{f2}
\end{figure}

Firstly, the depinning transition of the $2D$ driven RFIM is
measured carefully by the short time dynamics in Fig.~\ref{f1}(b)
with different driving filed $H$ and the disorder is fixed as
$\Delta = 1.5$. The transition point $H_c = 1.2933(2)$ is obtained
according to general scaling theory, which is more precise than $H_c
= 1.290(5)$ obtained through the stead state near the depinning
transition \cite{now98}. Then the exponent $\beta / \nu z =
0.217(2)$ is measured from the slope of the curve at the transition
point, according to Eq.~(\ref{equ90}). And the finite size effect is
also checked with the different lattice size $L = 256$ and $1024$.
Then in Fig.~\ref{f2}(a), the fluctuation function $F(t)$, defined
in Eq.~(\ref{equ60}), is plotted for the depinning transition and
its background, respectively. And basing on Eq.~(\ref{equ160}), the
exponent $1 / z = 0.749(5)$ and $1 / z_b = 0.677(3)$ are derived
from the slopes of the dash lines. Here $z$ is the dynamic exponent
for the depinning transition with $\Delta = 1.5, H_c = 1.2933$ and
$z_b$ is for the background with $\Delta = 0, H_c = 1.2933$. Then
the time evolution of the derivative $\partial_{\tau}\ln v(t,
\tau)|_{\tau=0}$ is displayed in Fig.~\ref{f2}(b) with open circles
and its slope $0.729(4)$ is measured. In order to fit the numerical
data well, a power law correlation is introduced, according to
Eq.~(\ref{equ170}). With Eq.~(\ref{equ100}), $1 / \nu z = 0.735(3)$
can be measured. Following, in Fig.~\ref{f3}(a), the roughness
function $\omega^2(t)$ is plotted with open dots and their slopes
are $1.701(6)$ for the depinning transition and $0.666(4)$ for the
background. Then the pure roughness function $D\omega^2(t)$, defined
in Eq.~(\ref{equ110}), is shown with pluses and the exponent $2\zeta
/ z = 1.717(5)$ is obtained according to Eq.~(\ref{equ130}), using
this power law correction form again. In the same way, the exponent
$2\zeta_b / z_b = 0.649(4)$ is also derived. Here $\zeta$ and
$\zeta_b$ denote the roughness exponent for the depinning transition
and the background, respectively.

    Combing the exponents $\beta / \nu z, 1 / \nu z,  1 /z $ and
$2\zeta / z$, one can calculate these critical exponents $\beta =
0.295(2), \nu = 1.02(1), z = 1.33(1)$ and $\zeta = 1.14(1)$
independent for the depinning transition. Then for its background,
$z_b = 1.50(1)$ and $\zeta_b = 0.487(5)$ are obtained. Finally, in
Fig.~\ref{f3}(b), the curves about the pure height correlation
function $DC(r,t)$ are plotted with open dots. For a sufficiently
large scale $r \gg \xi(t)$, e.g., $r = 256$, the exponent $2 \zeta /
z = 1.701(7)$, derived with the scaling analysis in
Eq.~(\ref{equ140}), is well consistent with $1.717(5)$ obtained in
Fig.~\ref{f3}(a). What's more, for a sufficiently small scale $r \ll
\xi(t)$, e.g., $r = 2$, the curve also obeys a power law behavior
and the exponent $2 (\zeta - \zeta_{loc}) / z = 0.597(4)$ is also
obtained. With $z = 1.33(1)$ at hand, one can calculate that
$2\zeta_{loc} = (1.701 - 0.597) \times 1.33 = 1.47$. Then in the
inset, the height correlation function $DC(r,\infty)$ vs. $r$ is
plotted for the depinning transition at the sufficiently larger
time, e.g., $t = 1024$.  And a special scaling form in
Eq.~(\ref{equ150}) is used to fit the data in the whole regime and
$2\zeta_{loca} = 1.46(3)$ is measured for the depinning transition,
which is also well consistent with the exponent $2\zeta_{loc} =
1.47$ obtained before.

    Finally, the critical exponents characterizing the velocity of
the domain interface, the roughness function and the height
correlation function are extracted, and all the results are
summarized in the last column of Table~\ref{t1}. For comparison,
some exponents, obtained by the driven QEW equation, the driven RFIM
in the stead state, the discrete dislocation dynamics at the
depinning transition and the forced-flow imbibition in columnar
geometries, are also listed here. As was mentioned in the
introduction, there are arguments in favor of the conjecture that
the motion of a domain wall in a random-field system can be
described by the driven QEW equation with quenched disorder
\cite{bru84}. While our findings doesn't support this conjecture,
because there are at least $10\%$ difference between these two
models. Hence, they are in the different universality classes. While
the theory values of these exponents for QEW have been derived by
the function renormalization group(FRG) with $z = 1.33, \zeta =
1.00$ for the one-loop order and $\beta = 0.31, \nu = 0.98, \zeta =
1.43$ for the two-loop order \cite{dou02}. With this, one can see
that some critical exponents in our work is well consistent with the
theory value. Then these two measurements of the roughness exponent
$\zeta$ by the pure roughness function $D\omega^2(t)$ and the pure
height correlation function $DC(r,t)$ are in good agreement. And
$\zeta = 1.13(1) > 1$ and $\zeta_{loc} = 0.743(6) < 1$ mean that the
interface growth process with intrinsic anomalous scaling and
spatial multiscaling takes place. It is quite different with the
interface growth process with superrough scaling and spatial single
scaling in QEW with $\zeta = 1.23 > 1$ and $\zeta_{loc} = 0.98
\approx 1$ . The dynamic equation QEW is used to describe a
one-dimensional driven elastic string in a two-dimensional
disordered medium. Then the position of the interface can be defined
uniquely and there are practically no overhangs, bubbles or
droplets. So the single spatial scaling behavior is easy to
understand. While overhangs is obvious in our work of RFIM, shown in
Fig.~\ref{f1}(a). Hence, we guess it is overhangs cause this spatial
multiscaling behavior, which is rarely in QEW. And additional
simulation for the background in our work also supports this
hypothesis. In our work, we confirm that the one-order transition
takes place at $H_c = \Delta$, for the pinning-depinning process
with $\Delta \leq 1$ \cite{rot99}. Hence, in our simulation of the
background without disorder, a steady velocity state is obtained
with the large driving field $H \gg H_c =0$. And a similar work has
also been performed in the driven elastic string in a disordered
medium \cite{due05}. To our surprise, the difference of the
roughness exponent $\zeta$ between these two models are neglectable
within the errors, different form that at the depinning transition.
And the overhangs vanish in the background with the finite velocity,
which can be observed in Fig.~\ref{f1}(a). Hence, we can believe
that overhangs lead to the main difference between this spin model
and QEW. And $\zeta = 0.5$ is expected, for the quenched disorder
acts effectively as a thermal noise at the largest scale with
$\zeta_{therm} = 1/2$ \cite{bus08,kol05,kol06a}.

  Then in the driven RFIM, the transition point $H_c$ and critical
exponents $\beta, \nu, z, \zeta$ can also be measured in the steady
state with finite-size scaling and finite-temperature scaling forms.
Based on the short-time dynamics scaling form, it is more convenient
and precise for the determination of both dynamic and static
exponents as well as transition point. Since one does not suffer
from critical slowing down. For instance, the transition point $H_c$
is reported as $1.290(5)$ in Ref. \cite{now98}. While this curve
with $H = 1.290$ visibly departure from the power law behavior,
shown in Fig.~\ref{f1}(b). It means that the transition point $H_c$
is distinctly above $1.290$. What's more, in this work, we can
investigate the roughness behavior carefully and systematically at
the depinning transition. While for the RFIM in the steady state, it
is too hard to measure the roughness exponent $\zeta$ and dynamic
exponent $z$. And in Table.\ref{t1}, these exponents with $+$ are
obtained in the critical avalanches process, instead of depinning
transition. Then the critical exponents of the depinning transition
is also obtained for the discrete dislocation dynamics, which are
quite agree with the result of our work.

Experimental investigation of domain wall motion in magnetic systems
takes place at finite temperatures. With the thermal activation, the
sharp depinning transition is clearly rounded. In the creep regime
with $H \ll H_c, T > 0$, the interface velocity does not vanish and
the scaling behavior has been investigated
\cite{kol06a,cha00,kol05}. And the local roughness exponent has been
measured experimentally, shown in Tab.~\ref{t2}. It is a litter
larger than $\zeta_{eq} = 2/3$, obtained in the equilibrium state
with $H = 0, T > 0$. Then in Refs.\cite{bus08,now98,rot01},
influence of temperature on the depinning transition of driven
interfaces has also been shown with $H = H_c, T > 0$ and the
roughness exponent $\zeta = 1.25$ is obtained, the same as the one
at the depinning transition \cite{bus08}. And in the
Ref.\cite{kol06a},  $\zeta = 1.26(1)$ can also be obtained, even at
the driving field $H$ well below $H_c$. While in these experiment,
$H$ is not so small. Especially in Ref.\cite{jos98}, $H = 17.7 kA/m$
is used, quite near the critical field $H_c \approx 40 kA/m$. And
these experimental result of the local roughness exponent are close
to $\zeta_{loc} = 0.735(5)$, measured in our work. Hence, a
crossover takes place from the equilibrium state to the depinning
states, as the driving filed is increased \cite{kol06a}. And in
Ref.\cite{kol05}, a similar crossover has been investigated as the
temperature is decreased.

\section{Conclusion}

   In summary, with Monte Carlo simulation we have investigated the
non-equilibrium critical dynamics of the two-dimensional driven
random-field Ising model for the depining transition and its
background at zero temperature. The dynamic scaling behavior is
carefully analyzed, and a dynamic roughening process is observed.
The transition point and critical exponents are totally listed in
Table~\ref{t1}. Based on the short-time dynamics, it is more
convenient and precise to measure these exponents than the work in
the steady state close to the depinning transition. The conjecture
that the motion of a domain wall in a random-field system can be
described by the equation EW with quenched disorder is in doubt
here. And intrinsic anomalous scaling and spatial multiscaling are
observed in the interface growth process of our work, quite
different from the superrough scaling and spatial single scaling in
QEW. When the overhangs vanish with the large driving field, the
difference between these two models disappears. Hence, we conjecture
that overhangs lead to the main difference between QEW and RFIM at
the depinning transition. What's more, the local roughness obtained
in this work is supported by the results of experiment, shown in
Tab.\ref{t2}. But the effect of overhangs should be studied further.
And it is also important to investigate the short-time dynamic
behavior at lower temperature in RFIM and/or the transition between
the relaxation, creep, slide, and switch regions with an oscillating
driving field $H = H_0\exp(i \omega t)$. Finally, the techniques
used in this paper can be also applied to similar dynamic systems,
such as xy model.

{\bf Acknowledgements:} This work was supported in part by NNSF
(China) under grant No. 10875102.


\bibliography{domain,zheng}

\begin{thebibliography}{10}

\bibitem{jan89}
{H.K. Janssen, B. Schaub, and B. Schmittmann}, Z. Phys. {\bf B73},  539
  (1989).

\bibitem{luo98}
{H.J. Luo, L. Sch\"ulke, and B. Zheng}, Phys. Rev. Lett. {\bf {81}},  180
  (1998).

\bibitem{zhe98}
B. Zheng, Int. J. Mod. Phys. {\bf B12},  1419  (1998), review article.

\bibitem{zhe99}
{B. Zheng, M. Schulz, and S. Trimper}, Phys. Rev. Lett. {\bf {82}},  1891
  (1999).

\bibitem{li95}
{Z.B. Li, L. {Sch\"ulke}, and B. Zheng}, Phys. Rev. Lett. {\bf {74}},  3396
  (1995).

\bibitem{zhe03a}
{B. Zheng, F. Ren, and H. Ren}, Phys. Rev. {\bf {E68}},  046120  (2003).

\bibitem{yin04}
{J.Q. Yin, B. Zheng, and S. Trimper}, Phys. Rev. {\bf E70},  056134  (2004).

\bibitem{yin05}
{J.Q. Yin, B. Zheng, and S. Trimper}, Phys. Rev. {\bf E72},  036122  (2005).

\bibitem{zho07}
{N.J. Zhou and B. Zheng}, Europhys. Lett. {\bf 78},  56001  (2007).

\bibitem{zho08}
{N.J. Zhou and B. Zheng}, Phys. Rev. {\bf E77},  051104  (2008).

\bibitem{lei07}
{X.W. Lei and B. Zheng}, Phys. Rev. {\bf E75},  040104  (2007).

\bibitem{sch00}
{L. Sch\"ulke and B. Zheng}, Phys. Rev. {\bf {E62}},  7482  (2000).

\bibitem{lin08}
{S.Z. Lin and B. Zheng}, Phys. Rev. {\bf E78},  011127  (2008).

\bibitem{lem99}
{S.G. Lemay, R.E. Thorne, Y. Li and J.D. Brock}, Phys. Rev. Lett. {\bf 83},
  2793  (1999).

\bibitem{fuc98}
{D.T. Fuchs, E. Zeldov, T. Tamegai, S. Ooi, M. Rappaport, and H. Shtrikman},
  Phys. Rev. Lett. {\bf 80},  4971  (1998).

\bibitem{dou08}
{A. Dourlat, V. Jeudy, A. Lema\^{i}tre, and C. Gourdon}, Phys. Rev. {\bf B78},
  161303(R)  (2008).

\bibitem{yam07}
{M. Yamanouchi, J. Ieda, F. Matsukura, S.E. Barnes, S. Maekawa, and H. Ohno},
  Science {\bf 317},  1726  (2007).

\bibitem{tyb02}
{T. Tybell, P. Paruch, T. Giamarchi, and J.-M. Triscone}, Phys. Rev. Lett. {\bf
  89},  097601  (2002).

\bibitem{par05}
{P. Paruch, T. Giamarchi, and J.-M. Triscone}, Phys. Rev. Lett. {\bf 94},
  197601  (2005).

\bibitem{mou04}
{S. Moulinet, A. Rosso, W. Krauth, and E. Rolley}, Phys. Rev. {\bf E69},
  035103(R)  (2004).

\bibitem{dou06}
{P.L. Doussal, K.J. Wiese, E. Raphael, and R. Golestanian}, Phys. Rev. Lett.
  {\bf 96},  015702  (2006).

\bibitem{he92}
{S.J. He, G.L.M.K.S. Kahanda, and P.Z. Wong}, Phys. Rev. Lett. {\bf 69},  3731
  (1992).

\bibitem{zap01}
{S. Zapperi and M. Zaiser}, Mater. Sci. Eng. {\bf A309},  348  (2001).

\bibitem{bak08}
{B. Bak\'o, D. Weygand, M. Samaras, W. Hoffelner, and M. Zaiser}, Phys. Rev.
  {\bf B78},  144104  (2008).

\bibitem{pon06}
{L. Ponson, D. Bonamy, and E. Bouchaud}, Phys. Rev. Lett. {\bf 96},  035506
  (2006).

\bibitem{rot99}
{L. Roters, A. Hucht, S. L\"ubeck, U. Nowak, and K.D. Usadel}, Phys. Rev. {\bf
  E60},  5202  (1999).

\bibitem{rot01}
{L. Roters, S. L\"ubeck, and K.D. Usadel}, Phys. Rev. {\bf E63},  026113
  (2001).

\bibitem{now98}
{U. Nowak and K.D. Usadel}, Europhys. Lett. {\bf 44},  634  (1998).

\bibitem{dou02}
{P.L. Doussal, K.J. Wiese, and P. Chauve}, Phys. Rev. {\bf B66},  174201
  (2002).

\bibitem{jos96}
{M. Jost and K.D. Usadel}, Phys. Rev. {\bf B54},  9314  (1996).

\bibitem{jos97}
{M. Jost and K.D. Usadel},  in {\em {Chaos and Fractals in Chemical
  Engineering}}, edited by {G. Biardi, M. Giona, and A.R. Giona} (World
  Scientific, Singapore, 1997).

\bibitem{due05}
{O. Duemmer and W. Krauth}, Phys. Rev. {\bf E71},  061601  (2005).

\bibitem{kol06}
{A.B. Kolton, A. Rosso, E.V. Albano, and T. Giamarchi}, Phys. Rev. {\bf B74},
  140201(R)  (2006).

\bibitem{kol06a}
{A.B. Kolton, A. Rosso, T. Giamarchi, and W. Krauth}, Phys. Rev. Lett. {\bf
  97},  057001  (2006).

\bibitem{bus08}
{S. Bustingorry, A.B. Kolton, and T. Giamarchi}, Europhys. Lett. {\bf 81},
  26005  (2008).

\bibitem{nat92}
{T. Nattermann, S. Stepanow, L.H. Tang, and H. Leschhorn}, J. Phys. II France
  {\bf 2},  1483  (1992).

\bibitem{jen95}
{H. J. Jensen}, J. Phys. A: Math. Gen. {\bf 28},  1861  (1995).

\bibitem{les93}
{H. Leschhorn and L.H. Tang}, Phys. Rev. Lett. {\bf 70},  2973  (1993).

\bibitem{tan92}
{L.H. Tang}, J. Stat. Phys. {\bf 67},  819  (1992).

\bibitem{ros01}
{A. Rosso and W. Krauth}, Phys. Rev. Lett. {\bf 87},  187002  (2001).

\bibitem{ros03}
{A. Rosso, A.K. Hartmann, and W. Krauth}, Phys. Rev. {\bf E67},  021602
  (2003).

\bibitem{goo04}
{T. Goodman and S. Teitel}, Phys. Rev. {\bf E69},  062105  (2004).

\bibitem{cha00}
{P. Chauve, T. Giamarchi, and P.L. Doussal}, Phys. Rev. {\bf B62},  6241
  (2000).

\bibitem{pan00}
{N.N. Pang and W.J. Tzeng}, Phys. Rev. {\bf E61},  3559  (2000).

\bibitem{pan00a}
{N.N. Pang and W.J. Tzeng}, Phys. Rev. {\bf E61},  3212  (2000).

\bibitem{pra08}
{M. Pradas, A.H. Machado, and M.A. Rodr\'{i}guez}, Phys. Rev. {\bf E77},
  056305  (2008).

\bibitem{aug06}
{M.A. Auger, L. V\'{a}zquez, R. Cuerno, M. Castro, M. Jergel, and O.
  S\'{a}nchez}, Phys. Rev. {\bf B73},  045436  (2006).

\bibitem{sai02}
{M. Saitou}, Phys. Rev. {\bf B66},  073416  (2002).

\bibitem{sai03}
{M. Saitou, K. Hamaguchi, and W. Oshikawa}, J. Electrochem. Soc. {\bf 150},
  C99  (2003).

\bibitem{lem98}
{S. Lemerle, J. Ferr\'{e}, C. Chappert, V. Mathet, T. Giamarchi, and P.L.
  Doussal}, Phys. Rev. Lett. {\bf 80},  849  (1998).

\bibitem{met07}
{P.J. Metaxas, J.P. Jamet, A. Mougin, M. Cormier, J. Ferr\'{e}, V. Baltz, B.
  Rodmacq, B. Dieny, and R.L. Stamps}, Phys. Rev. Lett. {\bf 99},  217208
  (2007).

\bibitem{kle07}
{W. Kleemann, J. Rhensius, O. Petracic, J. Ferr\'e, J.P. Jamet, and H. Bernas},
  Phys. Rev. Lett. {\bf 99},  097203  (2007).

\bibitem{bra05}
{T. Braun, W. Kleemann, J. Dec, and P.A. Thomas}, Phys. Rev. Lett. {\bf 94},
  117601  (2005).

\bibitem{jos98}
{M. Jost, J. Heimel and T. Kleinefeld}, Phys. Rev. {\bf B57},  5316  (1998).

\bibitem{nat01}
{T. Nattermann, V. Pokrovsky, and V.M. Vinokur}, Phys. Rev. Lett. {\bf 87},
  197005  (2001).

\bibitem{gla03}
{A. Glatz, T. Nattermann, and V. Pokrovsky}, Phys. Rev. Lett. {\bf 90},  047201
   (2003).

\bibitem{bru84}
{R. Bruinsma and G. Aeppli}, Phys. Rev. Lett. {\bf 52},  1547  (1984).

\bibitem{zhe01}
{G.P. Zheng and M. Li}, Phys. Rev. {\bf E63},  036122  (2001).

\bibitem{sep98}
{E.T. Sepp\"al\"a, V. Pet\"aj\"a, and M.J. Alava}, Phys. Rev. {\bf E58},  R5217
   (1998).

\bibitem{rod07}
{G.R. Rodr\'{i}guez, A.P. Junquera, M. V\'{e}lez, J.V. Anguita, J.I.
  Mart\'{i}n, H. Rubio, and J.M. Alameda}, J. Phys. D: Appl. Phys. {\bf 40},
  3051  (2007).

\bibitem{kol05}
{A.B. Kolton, A. Rosso, and T. Giamarchi}, Phys. Rev. Lett. {\bf 94},  047002
  (2005).

\bibitem{nol94}
{C.S. Nolle, B. Koiller, N. Martys, and M.O. Robbins}, Physica {\bf A205},  342
   (1994).

\bibitem{ama94}
{L.A.N Amaral, A.L Barab\'{a}si, and H.E. Stanley}, Phys. Rev. Lett. {\bf 73},
  62  (1994).

\bibitem{lop97}
{J.M. L\'{o}pez and M.A. Rodr\'{i}guez}, J. Phys. I France {\bf 7},  1191
  (1997).

\end{thebibliography}
\bibliographystyle{prsty}

\newpage

\begin{table}[h]\centering
\begin{tabular}[t]{r c c l c c}
\hline
\hline  & Exponent & QEW  & RFIM & Dislocation & This work\\
\hline
$v(t)$        &   $H_c$           &                                        &1.290(5)\cite{now98}                                         &                     &  1.2933(2)  \\
              &   $\beta$         & 0.33\cite{kol06};0.33(2)\cite{due05}   &0.35(4)\cite{now98};0.42(5)\cite{nol94};0.31(8)\cite{ama94}  & 0.30(5)\cite{bak08} &  0.295(2)   \\
              &   $\nu$           & 1.33\cite{kol06};1.29(5)\cite{due05}   &1.00(5)\cite{now98};1.33\cite{nol94}                         & 1.05(5)\cite{bak08} &  1.02(1)    \\
              &   $z $            & 1.5\cite{kol06};1.54(5)\cite{dou02}    &                                                             & 1.32(4)\cite{bak08} &  1.33(1)    \\
$\omega^2(t)$ &   $\zeta$         & 1.25\cite{kol06};1.26(1)\cite{due05}   &                                                             &                     &  1.14(1)    \\
$C(r,t)$      &   $\zeta$         & 1.2\cite{lop97};1.23(1)\cite{pan00a}    &                                                             & 0.98(3)\cite{bak08} &  1.13(1)    \\
              &   $\zeta_{loc}$   & 0.92\cite{lop97};0.98\cite{jos97}      &                                                             & 0.96(2)\cite{bak08} &  0.735(5)    \\ \\
$H \gg H_c$   &   $z_b$           & 1.5\cite{kol06}                        &                                                             &                     &  1.50(1)    \\
              &   $\zeta_b$       & 0.5\cite{due05}                        &0.5\cite{nol94}                                                             &                     &  0.487(5)   \\
\hline \hline
\end{tabular}
\caption{Summary of critical exponents are shown for the depinning
transition (upper sector) and its background(lower sector), obtained
with different techniques. Second column: result of numerical
simulation obtained for the driven Edwards-Wilkinson equation with
quenched disorder (QEW). Third column: result of numerical
simulation obtained for the driven random field Ising model (RFIM)
close to the transition point in the steady state. Forth column:
result of numerical simulation obtained for the discrete dislocation
model at the depinning transition. Then our work is shown in the
last column, and the result of numerical simulation is obtained for
the random field Ising model at the depinning transition with the
short-time critical dynamic.} \label{t1}
\end{table}

\newpage

\begin{table}[h]\centering
\begin{tabular}[t]{r @{\hspace{10mm}} *{5}{p{20mm}}}
\hline
\hline          &  This work   & Ref.\cite{lem98} & Ref.\cite{jos98} & Ref.\cite{met07}  & Ref.\cite{kle07}\\
$\zeta_{loc}$   &  0.735(5)    & 0.69(7)          & 0.78(1)          & 0.7(1)            & 0.6(1)  \\
\hline \hline
\end{tabular}
\caption{ The local roughness exponent for the depinning transition
in our work is shown here. In comparison, the result of experimental
investigation of domain wall motion in magnetic systems are also
listed in its creep regime. Refs.\cite{met07,lem98,kle07} work in
the ultrathin Pt/Co/Pt films and Ref.\cite{jos98} works in
Co$_{28}$Pt$_{72}$ alloy films. And an oscillating driving field
$H(t)$ is used in Ref.\cite{kle07}, instead of a dc driving field.
}\label{t2}
\end{table}

\newpage

\end{document}